\documentclass[prl, twocolumn, aps]{revtex4}
\usepackage{graphicx}
\usepackage{bm}
\usepackage{amssymb}
\usepackage{amsmath}
\usepackage{amsfonts}
\usepackage{amsopn} 
\DeclareMathOperator{\trace}{tr}

\newcommand* {\vekc}[1]{\ensuremath{\bm{\mathcal{#1}}}}

\newcommand {\bra} [1] {\langle #1 |}
\newcommand {\ket} [1] {| #1 \rangle}
\newcommand {\pd} [2] {\frac{\partial #1}{\partial #2}}

\begin{document}



\title{Generation of spin currents and spin densities in systems
with reduced symmetry}
\author{Dimitrie Culcer}
\author{R. Winkler}
\affiliation{Advanced Photon Source, Argonne National Laboratory,
Argonne, IL 60439.} \affiliation{Northern Illinois University, De
Kalb, IL 60115.}
\begin{abstract}
  We show that the spin-current response of a semiconductor crystal
  to an external electric field is considerably more complex than
  previously assumed. While in systems of high symmetry only the
  spin-Hall components are allowed, in systems of lower symmetry
  other non-spin-Hall components may be present. We argue that, when
  spin-orbit interactions are present only in the band structure, the
  distinction between intrinsic and extrinsic contributions to the
  spin current is not useful. We show that the generation of
  spin currents and that of spin densities in an electric field are
  closely related, and that our general theory provides a systematic
  way to distinguish between them in experiment. We discuss also the
  meaning of vertex corrections in systems with spin-orbit
  interactions.
\end{abstract}
\date{\today}
\maketitle

Ground-breaking work in past years has turned semiconductor spin
electronics into a richly rewarding field, both theoretically and
experimentally. The application of an electric field to a
semiconductor sample gives rise to a nonequilibrium spin current
\cite{dya71, wun05, awsch, sin04} as well as a steady-state spin
density in the bulk of the sample \cite{vor79, aro91, ede90}. The
latter was first observed several decades ago \cite{vor79}, while
the recent imaging \cite{cro05} and \emph{direct} measurements of
spin currents \cite{tin06, liu06}, together with the achievement of
the room-temperature spin-Hall effect \cite{ste06} have brought the
promise of spin-based electrical devices closer to fruition and
stimulated an enormous amount of research \cite{eng06}. Other
schemes for generating and detecting spin currents have been
implemented or proposed \cite{gan01}.

Aside from the specter of technological advances, the development of
electrical spin manipulation techniques has brought the fundamental
physics of spin transport under intense scrutiny \cite{shi06, ras04,
zha05, she05, wang06, han06, kha06, shy06}. Debate has focused on
definitions of spin currents \cite{shi06}, on whether spin currents
are transport or background currents \cite{ras04}, whether
scattering-independent or scattering-dependent contributions are
dominant \cite{eng06}, on the role played by spin Coulomb drag
\cite{han06}, and on spin accumulation at the boundary \cite{mal07}.
Whereas most studies to date have paid significant attention to
common semiconductors and asymmetric quantum wells \cite{dam04,
schl05, jin06, li04, ino04, sch02, dim05, ino03, mal07, eng05,
nik06, ble06, kro06, zar06}, recent developments call for an
in-depth investigation of neglected aspects.

In this article, we first discuss the relationship between spin
currents in a crystal and the symmetry of the underlying lattice.
Such an analysis has been enlightening in the context of
nonequilibrium spin densities excited by an electric field. If the
response of the spin density $\bm{s}$ to an electric field $\bm{E}$
is given by $s^\sigma = Q^{\sigma}_j E_j$, nonzero components for
the material-specific spin density response tensor $Q^{\sigma}_j$
are permitted only in gyrotropic crystals \cite{vor79}. Yet such an
analysis has not been done for spin transport. We therefore
determine the components of the spin-current response tensor allowed
by symmetry in an electric field, providing systematic proof that
spin currents in response to an electric field can be much more
complex than the spin-Hall effect \cite{nik04}. This result is
completely general and is not sensitive to the definition of the
spin current or to whether the electric field is constant or
time-dependent. The subsequent calculations confirm these
predictions by determining the corrections to the density matrix
present in an electric field and demonstrating their intimate
relationship with spin precession. We argue that, if spin-orbit
interactions are present only in the band structure, only one
contribution to the spin current exists, which appears intrinsic in
the weak momentum-scattering limit and extrinsic in the strong
momentum-scattering limit. We show that spin currents and bulk spin
densities in an electric field arise from linearly independent
contributions to the density matrix, and that certain setups can
distinguish between them and measure effects due solely to spin
currents. Our work considers realistic scattering potentials and is
very relevant to experiment, where the high symmetry often assumed
in theoretical approaches is usually not present.

The spin current operator is defined as
$\hat{\mathcal{J}}^\sigma_{i} = {\textstyle\frac{1}{2}}
(\hat{s}^\sigma \hat{v}_i + \hat{v}_i \hat{s}^\sigma )$, where
$\hat{s}^\sigma$ represents the spin component $\sigma$, and the
velocity operator $\hat{v}_i = 1/\hbar \, (\partial \hat H/\partial
k_i)$, with $\hat H$ the Hamiltonian. An alternative, more realistic
definition of the spin current has been proposed \cite{shi06},
according to which $\hat{\mathcal{J}}^\sigma_{i} = d/dt\, (\hat r_i
\hat s^\sigma)$. Yet we remark that from a symmetry point of view
these two definitions are equivalent so that the following symmetry
analysis applies to both definitions. The spin current
$\hat{\vekc{J}}$ is a second rank tensor that can be decomposed into
a pseudoscalar part, an antisymmetric (spin-Hall) part, and a
symmetric part. The pseudoscalar part is $\trace (\hat{\vekc{J}}) =
{\textstyle\frac{1}{3}} \, \hat{\bm s} \cdot \hat{\bm v}$ and
represents a spin flowing in the direction in which it is oriented.
The symmetric and antisymmetric parts are given, respectively, by
${\textstyle\frac{1}{2}} (\hat{s}^\sigma \hat{v}_i \pm
\hat{v}_\sigma \hat{s}^i )$.
The pseudoscalar and symmetric parts will be referred to as
\emph{non-spin-Hall} currents. Under the full orthogonal group
only the antisymmetric (spin-Hall) components of $\hat{\vekc{J}}$ are
allowed, indicating that these components are always permitted by
symmetry.

In general, the spin current response of a crystal to an electric
field $\bm{E}$ is characterized by a material tensor $\bm{T}$
defined by $\mathcal{J}^\sigma_{i} = T^\sigma_{ij} E_j$. For the 32
crystallographic point groups the symmetry analysis \cite{bir74} for
the tensor $\bm{T}$ is established by means of standard
compatibility relations \cite{kos63}. One is particularly interested
in those groups in which non-spin-Hall components may be present.
Our calculations show that the pseudoscalar part of the spin current
is only allowed by 13 point groups, while the symmetric part is
allowed in all systems except those with point groups $O$, $T_d$
(zinc blende), or $O_h$ (diamond). Lower symmetries, allowing
non-spin-Hall currents, are characteristic of systems of reduced
dimensionality.

It is well known that the generation of a spin density by an
electric field is the inverse of the circular photogalvanic effect
\cite{vor79, gan01}, while the spin-Hall effect also has an inverse
\cite{liu06}. Since spin densities induced by electric fields are
restricted to gyrotropic crystals \cite{vor79} the same restriction
applies to the circular photogalvanic effect. Similarly, the
symmetry analysis developed here is also applicable to the inverse
spin-Hall effect which needs to be complemented by an inverse
non-spin-Hall effect in systems with reduced symmetry.

In order to verify the proposition that non-spin-Hall currents may
exist in a variety of crystals, we discuss spin currents induced by
an electric field in spin-1/2 electron systems. For electrons the
effective Hamiltonian is written as $H = H_\mathrm{kin} +
H_\mathrm{so}$, where $H_\mathrm{kin} = \hbar^2 k^2/2m^*$ and
$H_\mathrm{so} = \frac{1}{2} \, \bm{\sigma} \cdot \bm{\Omega}$,
where $\bm{\Omega}$ is a momentum-dependent effective Zeeman field.
In the weak momentum-scattering regime we have $\varepsilon_F
\tau_p/\hbar \gg \Omega \, \tau_p/\hbar \gg 1$, with $\tau_p$ the
momentum relaxation time and $\varepsilon_F$ the Fermi energy,
whereas in the strong momentum-scattering regime $\varepsilon_F
\tau_p/\hbar \gg 1 \gg \Omega \, \tau_p/\hbar$.

The system is described by a density operator $\hat{\rho}$, which is
expanded in a basis of definite wave vector as
\begin{equation}
\label{rhodef} \hat \rho = \sum_{n,n'}\sum_{\bm{k}, \bm{k}'}
\rho_{nn'\bm{k} \bm{k}'} (t) \ket{\psi_{n\bm{k}}(t)}\bra{\psi_{n'\bm{k}'}(t)}.
\end{equation}
A constant uniform electric field $\bm{E}$ is included in the
crystal momentum through the vector potential $\bm{A}$ such that
$\bm{k} = \bm{q} + e \, \bm{A}/\hbar$, and the wave functions are
chosen to have the form $\ket{\psi_{n\bm{k}}(t)} = e^{i \bm{q} \cdot
\bm{r}} \ket{u_{n\bm{k}}(t)}$, where $\ket{u_{n\bm{k}}(t)}$ are
lattice-periodic functions that are \emph{not} assumed to be
eigenfunctions of the crystal Hamiltonian. In this basis the matrix
elements $\rho_{nn'\bm{k}\bm{k}'} (t) $ form the density matrix and
the impurity potential has matrix elements
$\mathcal{U}_{\bm{k}\bm{k}'} \, \openone +
\mathcal{V}_{\bm{k}\bm{k}'}$, where $\mathcal{V}_{\bm{k}\bm{k}'}$ is
the spin-dependent part.

The time evolution of the density operator is given by the quantum
Liouville equation, which allows us to derive rigorously the time
evolution of the part of the density matrix $\rho \equiv
\rho_{nn'\bm{k}\bm{k}}$ diagonal in wave vector. We subdivide $\rho
= \rho_{0} + \rho_{E}$, where $\rho_{0}$ is given by the Fermi-Dirac
distribution, and the correction $\rho_{E}$ is due to the electric
field $\bm{E}$. To first order in $\bm{E}$, $\rho_{E}$ satisfies
\begin{equation}\label{eq:boltze}
\pd{\rho_{E}}{t} + \frac{i}{\hbar}\, [H, \rho_{E}] + \hat{J}\,
(\rho_{E}) = \frac{e\bm{E}}{\hbar} \cdot \pd{\rho_{0}}{\bm{k}},
\end{equation}
where $\hat{J} \, (\rho_{E})$ is the collision integral to be given
below. The eigenvalues of $H$ are $\epsilon_\pm = \varepsilon_0 \pm
\Omega/2$, with $\varepsilon_0 = \hbar^2 k^2/2m^*$. The spin current
operator is simply $\hat{\mathcal{J}}^\sigma_{i} = \hbar k_i
s^\sigma/m^* + \frac{1}{4}\hbar \, \partial \Omega^\sigma/\partial
k_i$. In electron systems we usually have $H_\mathrm{so} \ll
H_\mathrm{kin}$, thus $\rho_{0} \approx f_{0} \, \openone +
H_\mathrm{so} \, \partial f_{0}/\partial \varepsilon_0$, where
$f_{0} (\varepsilon_0)$ is the Fermi-Dirac distribution. $\rho_{E}$
is divided into a scalar part and a spin-dependent part, $\rho_{E} =
f_E \openone + S_E$. To first order in
$H_\mathrm{so}/H_\mathrm{kin}$ the scattering term can be expressed
as $ \hat J \, (\rho_{E}) = (\hat{J}_0 + \hat{J}_s + \hat{J}_v)\,
(f_{E}) + \hat{J}_0\, (S_{E})$, where
\begin{subequations}
\begin{eqnarray}\label{eq:Fermi0}
\hat{J}_0 \, (f_{E}) & = &  \frac{2\pi n_i}{\hbar}
\int\!\frac{d^dk'}{(2\pi)^d}\, |\mathcal{U}_{\bm{k}\bm{k}'}|^2
\big( f_E - f'_E \big) \, \delta(\varepsilon_0 - \varepsilon_0'),
\hspace{1.5em} \\
\hat{J}_s \, (f_E) & = & \frac{\pi n_i}{2\hbar}\, \bm{\sigma}
\cdot\! \int\!\frac{d^dk'}{(2\pi)^d}\,
|\mathcal{U}_{\bm{k}\bm{k}'}|^2 \,(f_E - f'_E) \nonumber\\ [1ex] &
& \times \{ (\hat{\bm{\Omega}} + \hat{\bm{\Omega}}')\,
[\delta(\epsilon_+ - \epsilon'_+) - \delta(\epsilon_- -
\epsilon'_-)] \nonumber \\ [1ex] & & + (\hat{\bm{\Omega}} -
\hat{\bm{\Omega}}')\, [\delta(\epsilon_+ - \epsilon'_-) -
\delta(\epsilon_- - \epsilon'_+)]\}, \label{eq:Fermis} \\ [1ex]
\label{eq:Fermiv} \hat{J}_v \, (f_E) & = & \frac{2\pi n_i}{\hbar}
\int\!\frac{d^dk'}{(2\pi)^d}\,   Y_{\bm{k}\bm{k}'} \big( f_E -
f'_E \big) \, \delta(\varepsilon_0 - \varepsilon_0').
\end{eqnarray}
\end{subequations}
In the above $n_i$ is the impurity density, $d$ is the
dimensionality of the system, primed quantities denote functions of
$\bm{k}'$, $\hat{\bm{\Omega}}$ is a unit vector along $\bm{\Omega}$,
and $Y_{\bm{k}\bm{k}'} = \int d^dk''/(2\pi)^d\,
(\mathcal{U}_{\bm{k}\bm{k}''} \mathcal{V}_{\bm{k}''\bm{k}'} +
\mathcal{V}_{\bm{k}\bm{k}''} \mathcal{U}_{\bm{k}''\bm{k}'})$.
Equation (\ref{eq:Fermi0}) is the usual scalar scattering term, Eq.\
(\ref{eq:Fermis}) is due to band structure spin-orbit coupling, and
Eq.\ (\ref{eq:Fermiv}) is due to spin-orbit coupling in the
impurities. We will assume henceforth that band structure spin-orbit
interactions are much stronger than those due to impurities.

For $f_{E}$ we obtain the standard correction $ f_E = (e\,\bm{E}
\,\tau_p / \hbar) \partial f_0/\partial\bm{k} \, \openone$. As a
result, the effective source term that enters the equation for $S_E$
is $\Sigma_s - \hat{J}_s\, (f_E)$, where $\Sigma_s$ is the
spin-dependent part of $(e\bm{E}/\hbar) \partial\rho_{0} /
\partial\bm{k}$. In analogy with the customary Gram-Schmidt
orthogonalization of vectors, this effective source term can be
divided into two parts, $\Sigma_s - \hat{J}_s\, (f_E) = \Sigma_\|
+ \Sigma_\perp$, of which $\Sigma_\|$ commutes with the spin-orbit
Hamiltonian
\begin{equation}
\Sigma_\| = \frac{\trace \{[\Sigma_s - \hat{J}_s\,
(f_E)]\, H_\mathrm{so}\}}{\trace (H_\mathrm{so}^2)}\, H_\mathrm{so},
\end{equation}
while $\Sigma_\perp$ is the remainder. In matrix language
$\Sigma_\perp$ is \emph{orthogonal} to the Hamiltonian, thus $\trace
(\Sigma_\perp\, H_\mathrm{so}) = 0$. To find $\Sigma_\|$ and
$\Sigma_\perp$ we define projectors $P_\|$ and $P_\perp$ onto and
orthogonal to $H_\mathrm{so}$, respectively. Acting on the basis
matrices $\bm{\sigma}$, $P_\| \bm{\sigma} = 2\bm{\Omega}\,
H_\mathrm{so}/\Omega^2$, while $P_\perp \sigma_{x} = [(\Omega_y^2 +
\Omega_z^2)\,\sigma_x - \Omega_x\Omega_y\, \sigma_y -
\Omega_x\Omega_z\, \sigma_z]/\Omega_k^2$, and the remaining terms
are obtained by cyclic permutations.

$S_E$ can likewise be divided into two linearly independent
contributions, $S_\|$ and $S_\perp$, the former of which commutes
with $H_\mathrm{so}$ while the latter is orthogonal to it. It is
helpful to think of $S_\|$ as the distribution of \emph{conserved}
spins parallel to $\bm{\Omega}$ and of $S_\perp$ as the distribution
of \emph{precessing} spins perpendicular to it. $S_\|$ satisfies
\begin{equation}
\pd{S_\|}{t} + P_\|[\hat{J}_0 \, (S_E)] = \Sigma_\| .
\end{equation}
This equation can be solved iteratively for any scattering. To solve
the equation for $S_\perp$, on the other hand, one needs to expand
$S_\perp$ in the strength of the scattering potential
$|\mathcal{U}|^2$, as $S_\perp = S_{\perp 0} + S_{\perp 1} +
O(|\mathcal{U}|^4)$. (It is easily shown that the first term in the
expansion must be zeroth order in $|\mathcal{U}|^2$, while $S_\|$
starts at order $-1$.) The equations for these contributions are
\begin{subequations}
\begin{eqnarray}
\pd{S_{\perp 0}}{t} + \frac{i}{\hbar}\, [H_\mathrm{so}, S_{\perp
0}] & = & \Sigma_\perp - P_\perp [\hat{J}_0 \, (S_\|)],
\label{eq:Sperp01a} \\ [1ex] \pd{S_{\perp 1}}{t} + \frac{i}{\hbar}
\, [H_\mathrm{so}, S_{\perp 1}] & = & P_\perp [\hat{J}_0 \,
(S_{\perp 0})] . \hspace{1em}
\end{eqnarray}
\end{subequations}
It follows that if $\Sigma_\perp - P_\perp [\hat{J}_0 \, (S_\|)]$
vanishes in Eq.\ (\ref{eq:Sperp01a}), as it does for the linear
Rashba model, then all the contributions to $S_\perp$ vanish. In
general, the equation for $S_{\perp 0}$ is also solved iteratively
for any scattering, but a closed-form solution for $S_E$ is not
always possible. An enlightening closed-form solution is, however,
possible for short-range impurities, where $\hat{J}_0\, (S_E) = (S_E
- \bar{S}_E)/\tau_p$, with the bar denoting averaging over
directions in $k$-space and $\tau_p = \hbar^3 / (m^*
|\mathcal{U}|^2)$. The series of equations of increasing order in
$|\mathcal{U}|^2$ give two geometric progressions that sum to
\begin{subequations}\label{eq:S}
\begin{eqnarray}\label{eq:Sparallel}
S_\| & = & \Sigma_\| \, \tau_p + P_\| \, (1 - \bar{P}_\|)^{-1}
\bar{\Sigma}_\| \, \tau_p, \\ [1ex]
\label{eq:Sperp} S_\perp & = & - \frac{\Sigma_\perp \,\tau_p +
P_\perp\, \bar{S}_\|}{1 + \Omega^2 \tau_p^2/\hbar^2} +
\frac{\bm{\Omega} \times (\bm{\Sigma}_\perp \tau_p + P_\perp\,
\bar{\bm{S}}_\| ) \cdot \bm{\sigma}\, \tau_p}{2\hbar(1 + \Omega^2
\tau_p^2/\hbar^2)}. \hspace{2.0em}
\end{eqnarray}
\end{subequations}
In the general case considered below the complex expressions for
$S_E$, $\Sigma_\|$ and $\Sigma_\perp$ will not be given explicitly.

Our analysis clarifies the relation between steady-state spin
currents and spin densities that appear when an electric field is
applied to a semiconductor. Since the spin operator is even in
$\bm{k}$ and the spin current operator is odd in $\bm{k}$, it
emerges, after evaluating $S_\|$ and $S_\perp$, that the RHS of Eq.\
(\ref{eq:Sparallel}) is responsible for steady-state spin densities
\cite{vor79, aro91, ede90}, while the second term on the RHS of Eq.\
(\ref{eq:Sperp}) gives rise to spin currents \cite{dya71, wun05,
awsch, sin04}. [The first term on the RHS of Eq.\ (\ref{eq:Sperp})
vanishes in both the weak and the strong momentum scattering
limits.] We conclude that spin densities arise from $S_\|$, the
distribution of conserved spin while spin currents arise from
$S_\perp$, the distribution of precessing spin. Thus nonequilibrium
spin currents are due to spin precession (as first demonstrated in
\cite{sin04}) while nonequilibrium spin densities \cite{vor79,
aro91, ede90} are due to the \emph{absence} of spin precession. The
physical picture for the latter mechanism is as follows. Each spin
on the Fermi surface precesses about an effective field ${\bm
\Omega} ({\bm k})$ and the spin component parallel to ${\bm \Omega}
({\bm k})$ is preserved. In equilibrium the average of the conserved
components is zero, but when an electric field is applied the Fermi
surface is shifted and the average of the conserved spin components
may be nonzero. This intuitive physical argument, to our knowledge
absent from the literature, explains why the nonequilibrium spin
density $\propto \tau_p^{-1}$ and \emph{requires} scattering to
balance the drift of the Fermi surface. We note that, on the other
hand, spin currents contain only terms $\propto \tau_p^{2n}$ with
$n=0, 1, 2, \ldots$.

It can be seen from Eq.\ (\ref{eq:Sperp}) that there is only
\emph{one} contribution to the spin current, which in the weak
momentum-scattering (intrinsic) limit is independent of $\tau_p$ and
in the strong momentum-scattering (extrinsic) limit is $\propto
\tau_p^2$. It was also noted that, if the RHS of Eq.\
(\ref{eq:Sperp01a}) vanishes, then $S_\perp$ vanishes to all orders.
We conclude that, if spin-orbit interactions exist only in the band
structure, the distinction between intrinsic and extrinsic spin
contributions to the spin current is not useful.

In calculations of spin currents $\bm{\mathcal{J}}$ based on the
Green's functions formalism vertex corrections play an important
role \cite{sch02, ino04, dim05}. The above analysis suggests that
contributions due to two processes are contained in these vertex
corrections. Scattering renormalizes the driving term for $S_E$ to
$\Sigma_s - \hat{J}_s\, (f_E)$, and it mixes the conserved and
precessing spin distributions, as in Eq.\ (\ref{eq:Sperp01a}). This
implies that vertex corrections to $\bm{\mathcal{J}}$ contain the
influence of a steady-state spin density on $\bm{\mathcal{J}}$. As
this spin density occurs only in gyrotropic materials \cite{vor79},
we expect vertex corrections to spin currents to vanish in
non-gyrotropic materials.

Engel \textit{et~al.} \cite{eng05} showed that, when band structure
spin-orbit interactions are negligible, spin currents arise from
skew scattering. A comparison of our results with those of Ref.\
\cite{eng05} shows that scattering can give rise to spin currents of
qualitatively different forms depending on whether spin-orbit
interactions are present in the band structure or not. In the
absence of band structure spin-orbit interactions, scattering
processes involve scattering of conserved spins, but in the presence
of band structure spin-orbit interactions, scattering processes
involve the scattering of precessing spins. A general analysis of
band structure spin-orbit interactions and skew scattering on the
same footing remains to be undertaken.

For known cases, our theory agrees with previous work. 2D
Hamiltonians with spin-orbit coupling linear in ${\bm k}$ give zero
spin current for any scalar scattering potential \cite{sug06,
dim05}, including short-range \cite{sch02, ino04, ino03, dim05,
kha06}, and small-angle scattering \cite{kha06, shy06}. For
Hamiltonians characterized by one Fourier component $N$ \cite{shy06}
the spin current $\propto N$. In 3D the \emph{correction} to the
spin current due to $\hat{J}_s\, (f_{E})$ vanishes for the
$k^3$-Dresselhaus model and short-range impurities. Our results also
agree with previous calculations of spin generation \cite{ede90}.

As a specific example, we investigate the spin current response
tensor $\bm{T}$ in systems of low symmetry. A strong justification
for this choice is that experiment often studies low-symmetry
structures whereas theory is often done for high-symmetry models.
To reduce the symmetry of a system one can either apply strain or
lower its dimensionality. In strained bulk samples, non-spin-Hall
currents are expected to be proportional to the strain tensor and
therefore small. We concentrate thus on quantum wells and,
considering for definiteness a symmetric 150-{\AA} wide GaAs well
grown along [113]. The axes of the coordinate system are
$x=[33\bar{2}]$, $y = [1\bar{1}0]$ and $z = [113]$. Spin-orbit
interactions are described by $k$-linear and $k^3$-Dresselhaus
terms \cite{win03} and impurity scattering by a screened Coulomb
potential, with the Thomas-Fermi wave vector $k_0 = 4k_F$ at
carrier density $n = 1.6\times 10^{11}$~cm$^{-2}$. We do not
consider skew scattering, which is important when band structure
spin-orbit interactions are weak \cite{eng05, ste06}. In units of
$e/(8\pi)$, we obtain $T^x_{xx} = 0.398$, $T^x_{yy} = 0.12$,
$T^z_{yx} = 0.172$, and $T^z_{xy} = -0.414$, which shows that for
realistic scattering in a system of low symmetry many components
of $\bm{T}$ are nonzero, excepting, of course, transport
perpendicular to the plane of the structure. At the boundary one
must study also the spin current according to the alternative
definition \cite{shi06}. Previous work \cite{shi06} has shown this
to be of the same order of magnitude as the conventional spin
current, with occasional sign differences. As our symmetry
analysis holds for both definitions of the spin current, we expect
the results to conform to the pattern found above.

Since several spin components can flow in the same direction and the
same spin component can flow in different directions, an interesting
experiment could detect spin currents. Taking the QW along [113]
considered above, an electric field applied along $x$ will produce a
nonequilibrium spin density, a spin-Hall current and a longitudinal
spin current composed of spin-$x$ only. If the QW is joined to a
material in which no nonequilibrium spin density is generated, then
a Kerr rotation \cite{awsch} or magnetic circular dichroism
\cite{sra06} experiment will give a nonzero signal exclusively due
to the injected spin-$x$ current.

In summary, we have shown that in systems with reduced symmetry spin
currents are not restricted to the spin-Hall effect, and that this
fact can help one distinguish experimentally between
electrically-induced spin densities and spin currents. We have
demonstrated in addition that spin currents in an electric field are
associated with spin precession, whereas spin densities are
associated with the absence of spin precession.

We acknowledge enlightening discussions with E.~Rashba, H.~A.~Engel,
Q.~Niu, A.~H.~MacDonald, D.~Xiao, E.~M.~Hankiewicz, G.~Vignale, and
J.~Sinova. The research at Argonne National Laboratory was supported
by the US Department of Energy, Office of Science, Office of Basic
Energy Sciences, under Contract No. DE-AC02-06CH11357.

\end{document}